\documentclass[fleqn,twoside]{article}
\usepackage{espcrc2}

\usepackage{mathrsfs}
\usepackage{bm}
\usepackage{stmaryrd}
\usepackage{slashed}
\usepackage{amsfonts}

\usepackage{graphicx}
\usepackage[figuresright]{rotating}

\newcommand{\AmS}{{\protect\the\textfont2
  A\kern-.1667em\lower.5ex\hbox{M}\kern-.125emS}}

\title{Towards neutrino transport with flavor mixing in supernovae: the Liouville operator}

\author{C. Y. Cardall\address{Physics Division, Oak Ridge National Laboratory, \\
      Oak Ridge, TN 37831-6354, United States of America}%
      \address{Department of Physics and Astronomy, University of Tennessee, \\
	Knoxville, TN 37996-1200, United States of America}%
        \thanks{Oak Ridge National Laboratory is managed by UT-Battelle, LLC, for the DoE under contract DE-AC05-00OR22725.}}
       
\begin{document}

\begin{abstract}
The calculation of neutrino decoupling from nuclear matter requires a transport formalism capable of handling both collisions and flavor mixing. The first steps towards such a formalism are the construction of neutrino and antineutrino `distribution matrices,' and a determination of the Liouville equations they satisfy in the noninteracting case. These steps are accomplished through study of a Wigner-transformed `density function,' the mean value of paired neutrino quantum field operators.
\vspace{1pc}
\end{abstract}

\maketitle



A massive star supports itself against gravity through the energy generated by a succession of nuclear fusion reactions; but once the core burns to iron-group elements it can extract no further energy from fusion, and when electron degeneracy can no longer support it, catastrophic collapse results.

The halt of collapse results in a shock wave that will eventually disrupt the entire star.
The inner part of the core collapses subsonically until it reaches nuclear density or so; but the outer part of the core collapses supersonically, and it doesn't receive word that the collapse of the inner core has halted until it slams into the inner core and forms a shock wave. The shock moves outwards before  stalling due to energy losses from neutrino emission and the breakup of heavy nuclei falling through the shock.

The mechanism of shock revival is still unknown, but simulations exploring the mechanism have to take account of the neutrino heating and cooling rates behind the shock. These rates depend on the energy and angle distributions of neutrinos, which must be tracked as the neutrinos decouple from the newly-born neutron star.

Microscopic quantum mechanical effects can be included if the tracking of neutrino energy and angle distributions is handled with classical transport.
 A classical distribution function $f(t,{\bf x},{\bf p})$ gives the differential number of neutrinos with momenta within $d^3{\bf p}$ of ${\bf p}$ at positions within $d^3{\bf x}$ of ${\bf x}$:
\begin{equation}
dN = f(t,{\bf x},{\bf p})\; \frac{g\, d^3{\bf p}}{(2\pi)^3} \; d^3{\bf x}. \label{classicalDistribution}
\end{equation}
 (For neutrinos the spin degeneracy $g$ is 1). The evolution of $f$ is given by the Boltzmann equation
 \begin{equation}
p^\mu \frac{\partial f}{\partial x^\mu} -{\Gamma^i}_{\nu\rho}\,p^\nu p^\rho\,\frac{\partial f}{\partial p^{i}}= C(f). \label{boltzmann}
\end{equation}
 The left-hand side is the Liouville operator. It is closely related to the geodesic equations describing classical particle trajectories in phase space. The first term gives the spacetime dependence, and the second term includes (for instance) particle redshifts and trajectory bending due to general relativity. The collision integral on the right-hand side describes particle interactions that add or remove particles from a particular trajectory. Rates computed with (for instance) quantum field theory can be used here, so that inherently quantum mechanical---but microscopic---phenomena such as particle creation and annihilation, and phase space blocking, can be included.

However, neutrino flavor mixing is a macroscopic quantum effect that cannot be treated with classical transport, which refers only to particle number and not amplitudes or phases.

What about the evolution of quantum occupation numbers rather than classical transport? An occupation number $n(t,{\bf q})$ gives the average number of particles occupying momentum eigenstate ${\bf q}$ in the (effectively infinite) quantization volume $V$. This is most naturally useful in the case of spatial homogeneity. The time derivative $dn/dt$ is given by quantum transition rates.

Noting that neutrino interaction rates have Ôoff-diagonalÕ elements when written in terms of `physical states' (or mass eigenstates), Dolgov \cite{Dolgov1981Neutrinos-in-th} expanded the occupation number to a matrix $\rho(t,|{\bf q}|)$ in flavor/mass space in a study of neutrinos in the (homogeneous and isotropic) early universe.

Keep in mind the differences between classical distribution functions $f(t,{\bf x},{\bf p})$ and quantum occupation numbers $n(t,{\bf q})$. The momentum of a classical particle is here labeled by ${\bf p}$, while a momentum eigenstate is labeled by ${\bf q}$. In the classical limit of a quantum description, ${\bf p}$ might be the centroid of a wave packet superposition of eigenstates ${\bf q}$.

However, in evolving the neutrino occupation matrix in the context of the early universe, the classical Liouville operator was employed to account for the redshift associated with cosmic expansion. That is, Dolgov tacitly assumed the substitution
\begin{equation}
\frac{d\rho(t,|{\bf q}|)}{dt} \rightarrow \frac{\partial\rho(t,|{\bf p}|)}{\partial t}- H |{\bf p}| \frac{\partial\rho(t,|{\bf p}|)}{\partial |{\bf p}|}, \label{replacement}
\end{equation}
in which the time derivative of the quantum occupation matrix goes over to the Liouville operator acting on a distribution matrix whose momentum dependence has its classical meaning.

Unlike the early universe case, neutrino transport in supernovae requires treatment of spatial inhomogeneity. The `Liouville replacement' of Eq. (\ref{replacement}) seems natural enough in the homogeneous case, but the introduction of spatial dependence should give one pause, because of the quantum mechanical incompatibility of position and momentum. Nevertheless, in the limit in which the neutrino trajectories are classical and only the flavor evolution is quantum mechanical, one expects the same terms as in the classical Liouville operator to emerge, including the spatial derivative term.

The Liouville operators for neutrino and antineutrino distribution matrices also contain effects of flavor mixing. In flat spacetime, and in the absence of interactions, the distribution matrices are governed by
\begin{eqnarray}
p^\mu \frac{\partial}{\partial x^\mu} \rho(t,{\bf x},{\bf p}) + \frac{{\rm i}}{2} \left[  \Delta, \rho(t,{\bf x},{\bf p}) \right] &=& 0, \label{neutrinoLiouville2}\\
p^\mu \frac{\partial}{\partial x^\mu} \bar\rho(t,{\bf x},{\bf p}) -  \frac{{\rm i}}{2} \left[ \Delta, \bar\rho(t,{\bf x},{\bf p}) \right] &=& 0. \label{antineutrinoLiouville2}
\end{eqnarray}
Species indices are suppressed. Only squared mass differences affect observable oscillation phenomena, and so a squared mass matrix $M^2$ in the commutator has been replaced by
\begin{equation}
\Delta = M^2 - \frac{1}{3}\mathrm{Tr}\left(M^2\right).
\end{equation}
The commutator enters with opposite sign in the case of antineutrinos if one wants the neutrino and antineutrino distribution matrices to transform in the same way under a flavor/mass basis change. 

How can the Liouville equations of Eqs. (\ref{neutrinoLiouville2}) and (\ref{antineutrinoLiouville2})---in particular, their spatial dependence---be derived rather than just assumed? 

One way is through a covariant version of the familiar simple model of flavor mixing \cite{Cardall2008Liouville-equat}. In this version a Schr\"odinger equation for a neutrino state is written down that involves not evolution in time or space alone, but along a worldline with affine parameter $\lambda$. One can build a density operator out of the neutrino states; the Liouville equation then follows from application of the Schr\"odinger equation to the kets and bras out of which the density operator is constructed, together with the relation $d/d\lambda = p^\mu \partial/\partial x^\mu$. 

But we will ultimately want a formalism in which collisions can be included, and as a first step towards this, the Liouville equations can be constructed from the equations of motion for a quantum `density function.' the mean value of paired neutrino quantum field operators \cite{Cardall2008Liouville-equat}.

Define a density function $\Gamma^{\ell m}_{ij}(y,z)$ as the mean value of a pair of normal-ordered neutrino quantum field operators:
\begin{equation}
{\rm i}\,\Gamma^{\ell m}_{ij}(y,z) = \left\langle N \nu_i^\ell(y)\, \bar \nu_j^m(z) \right\rangle. \label{densityFunction}
\end{equation}
It depends on spacetime positions $y$ and $z$, and has mass indices $i$ and $j$ and spinor indices $\ell$ and $m$. The $N$ denotes normal ordering. Writing the quantum field operator as $\nu(y) = A(y) + B(y)$ in terms of its positive- and negative-frequency parts $A(y)$ and $B(y)$, the density operator becomes
\begin{equation}
{\rm i}\,\Gamma^{\ell m}_{ij}(y,z) \!=\! -\left\langle \bar A_j^m(z) A_i^\ell(y) \right\rangle + \left\langle B_i^\ell(y) \bar B_j^m(z) \right\rangle, \label{densityFunction2}
\end{equation}
in which we have separate neutrino and antineutrino terms (rapidly oscillating cross terms not relevant to the macroscopic limit have been dropped).

The mixed representation of the density function begins to look more like a classical distribution function. Introduce the average position variable $x = (y+z)/2$ and the spatial difference variable $\Xi = y-z$. The mixed representation is obtained with a Wigner transformation, which is a Fourier transformation with respect to the difference variable: 
\begin{equation}
\mathcal{G}^{\ell m}_{ij}(x, P) = \int d^4 \Xi \; \mathrm{e}^{{\rm i} P\cdot \Xi} \;  \Gamma^{\ell m}_{ij}\left(x+\frac{\Xi}{2}, x - \frac{\Xi}{2}\right). \label{neutrinoWigner}
\end{equation}
Separate neutrino and antineutrino distributions $G^{\ell m}_{ij}(x, p)$ and $\bar G^{\ell m}_{ij}(x, p)$ are obtained from $\mathcal{G}^{\ell m}_{ij}(x, P)$ by projecting out its positive- and negative-frequency parts.

In general circumstances the mixed representation provides only complementary position or momentum distributions, when integrated over momentum space or the volume of the system respectively. This is in accordance with the quantum mechanical incompatibility of position and momentum.

However, a joint position/momentum distribution does emerge under classical conditions. This can be shown by explicitly evaluating the mixed representation with respect to a state of $N$ uncorrelated wave packets \cite{Cardall2008Liouville-equat}. After calculation, and arguments about Ôcoarse grainingÕ, and application of classicality conditions of weak inhomogeneity and slow time variation, a neutrino distribution matrix built on classical trajectories emerges:
\begin{eqnarray}
\rho_{ij}(t,{\bf x},{\bf p}) \!\!\!\!&=&\!\!\!\!
\sum_{a=1}^N (2\pi)^3\, \delta^3({\bf x} - {\bf x}_a(t))\; \delta^3({\bf p} - {\bf p}_a) \nonumber \\
& & \times \; \mathrm{e}^{-\mathrm{i}(E_{{\bf p}_a,i}-E_{{\bf p}_a,j})t} \,U_{\alpha_a j}\, U_{\alpha_a i}^*. \label{classicalNeutrinoDistribution}
\end{eqnarray}
(The antineutrino distribution matrix is the same except for a sign reversal in the exponential.)
Here ${\bf x}_a(t)$ is the classical trajectory followed by the $a$th neutrino, whose wave packet centroid is ${\bf p}_a$. This expression corresponds to a system of neutrinos born in flavors $\alpha_a$ at $t = 0$; the energies corresponding to the superposed mass eigenstate wave packet centroids are $E_{{\bf p}_a,i}$. The transformation to a flavor-basis distribution matrix follows directly from Eq. (\ref{densityFunction}) and the relation $\nu_\alpha(y) = U_{\alpha i} \, \nu_i(y)$ between flavor and mass fields. The diagonal elements of the flavor-basis distribution matrix are just as expected for a collection of neutrinos whose flavor content varies along classical trajectories according to the familiar vacuum oscillation probabilities.

The Liouville equations follow from the difference of Klein-Gordon equations
\begin{equation}
\left(\boxempty_y - \boxempty_z\right)  \Gamma^{\ell m}(y,z) + \left[ \Delta,  \Gamma^{\ell m}(y,z) \right] = 0 \label{kleinGordonDifference}
\end{equation}
obeyed by the density function. This can be seen by noting that $\boxempty_y - \boxempty_z = 2\, \frac{\partial}{\partial \Xi} \cdot \frac{\partial}{\partial x}$ in the coordinates of the mixed representation: when Eq. (\ref{kleinGordonDifference}) is transformed by the inverse of Eq. (\ref{neutrinoWigner}) and its positive- and negative-frequency parts are projected out, Eqs. (\ref{neutrinoLiouville2}) and (\ref{antineutrinoLiouville2}) are the result.

The calculations outlined here serve as a first step towards the handling of neutrino interactions with a diagrammatic approach based on a nonequilibrium Green's function \cite{Lifshitz1981Physical-Kineti}. In addition to a Green's function, this approach---which is sometimes called `Keldysh theory'---involves the density function considered in this paper and two other types of field operator pairings.

\def\prd{Phys. Rev. D }




\end{document}